\documentclass[runningheads]{llncs}

\usepackage[most]{tcolorbox}
\usepackage{multirow}
\usepackage{enumitem}
\usepackage{xcolor}
\usepackage{color}
\usepackage{xparse}    
\usepackage{url}
\usepackage{booktabs}
\usepackage{comment}
\usepackage{cite}

\newlist{rqlist}{description}{1}
\setlist[rqlist]{%
  font=\bfseries,      
  style=nextline,      
  labelsep=0.6em,      
  leftmargin=!,        
  widest={(RQ3)},      
  itemsep=.6ex, topsep=.6ex
}

\newcommand{\RQone}{How effective is ThinkLog compared with existing approaches?}
\newcommand{\RQtwo}{What is the impact of the length and number of reasoning shots on the performance of ThinkLog?}
\newcommand{\RQthree}{What is the impact of different LLMs used for reasoning generation on the performance of ThinkLog?}
\newcommand{\RQfour}{How well does ThinkLog generalize to unseen code?}

\NewDocumentCommand{\todo}{o}{
  \fcolorbox{red}{gray!25}{\small\textcolor{red}{\tt{TODO\IfNoValueTF{#1}{}{: #1}}}}
}

\usepackage{graphicx}
\usepackage{multirow}
\usepackage{xcolor}
\usepackage{comment}
\usepackage{balance}
\usepackage{bbding}

\begin{document}

\title{ThinkLog: Leveraging Reasoning for Log Statement Generation}

\author{Kazuki Kusama\inst{1}\orcidID{0009-0001-6623-3988}\Envelope \and
Honglin Shu\inst{1}\orcidID{0009-0005-7311-7060} \and
Masanari Kondo\inst{1}\orcidID{0000-0002-6317-7001} \and
Tao Xiao\inst{1}\orcidID{0000-0003-4070-585X} \and
Yasutaka Kamei\inst{1}\orcidID{0000-0002-7058-1045}}

\authorrunning{K. Kusama et al.}

\institute{Kyushu University, Fukuoka, Japan\\
\email{\{kusama,shu\}@posl.ait.kyushu-u.ac.jp}\\
\email{\{kondo,xiao,kamei\}@ait.kyushu-u.ac.jp}}

\maketitle

\begin{abstract}

Runtime logs are an important source of information that supports software maintenance.
To obtain useful logs, developers spend significant effort identifying appropriate log locations, assigning correct severity levels, and writing concise yet informative messages.
Therefore, end-to-end automated log statement generation can help reduce this burden, and prior work has proposed many methods for this task.
However, existing methods still exhibit limited accuracy.
To address this problem, we propose \textit{ThinkLog}, an LLM-based end-to-end log statement generation method.
The core idea of ThinkLog is to incorporate \textit{reasoning} that helps LLMs make decisions about log insertion, severity level assignment, and message generation, thereby improving log statement generation accuracy.
ThinkLog injects reasoning into prompts as few-shot examples and guides LLMs to generate appropriate log statements.
Evaluated on 9,619 Java methods extracted from public GitHub repositories, ThinkLog achieves 20.55\% log statement generation accuracy,  representing a 15.4\% improvement over the best existing method.
Moreover, these improvements were achieved at approximately 50\% of the inference cost (USD) compared to the best existing method.
These results show that leveraging reasoning is an effective and cost-efficient way to improve the accuracy of end-to-end log statement generation.
\end{abstract}


\section{Introduction} \label{sec:introduction}
As software systems grow in size and complexity, runtime logs serve as an important source of information for developers during maintenance activities (e.g., anomaly detection)~\cite{Chen2021ACMCS,He2021ACMCS}.
To ensure that such logs provide useful information, developers must carefully design them by selecting appropriate locations, assigning suitable severity levels, and generating log messages with sufficient contextual information.
This design process is inherently difficult and demands substantial time and effort~\cite{Fu2014ICSE,Zhu2015ICSE,Li2021TSE}.

To support log design, prior studies have actively explored automated \textit{log statement generation}~\cite{Guang2023TSE}.
Early approaches decomposed the log statement generation process into separate tasks (e.g., identifying where to insert log statements, determining appropriate severity levels, and generating log messages) and developed automated methods for each task~\cite{Li2021ASE,Liqun2022ISSTA}.
Beyond these task-specific methods, end-to-end methods have also been explored that integrate these tasks to fully automate log statement generation~\cite{Mastropaolo2022ICSE,Xie2024ISSTA,Xu2024ICSE}.
For example, LANCE~\cite{Mastropaolo2022ICSE} represents the first end-to-end approach, built on T5~\cite{T5}, a large language model (LLM).

Despite leveraging LLMs, end-to-end log generation remains prone to inaccuracy; for instance, FastLog~\cite{Xie2024ISSTA} achieves only $\approx$16\% accuracy. This limitation stems from neglecting the developer's intent. 
As Gu et al.~\cite{Guang2023TSE} highlight, while the \textit{Where} and \textit{What} are well-studied, the \textit{Why} (rationale) is frequently overlooked. Existing methods fail to explicitly model this reasoning, forcing LLMs to rely on surface-level patterns rather than logical inference~\cite{zhao2025chain}. 
This superficial processing precipitates failures, including variable hallucinations and misinterpretations of the code's state.

Providing LLMs with task-specific reasoning processes is a practical way to incorporate this missing why into LLM-based software engineering tasks~\cite{Yin2024ISSTA,Li2025TOSEM,Yang2026ICSE}. 
For example, in automated program repair, Yin et al.~\cite{Yin2024ISSTA} proposed ThinkRepair. This approach first prepares multiple chains of thought that capture the reasoning processes involved in bug fixing, such as analyzing the cause of a bug and determining a repair strategy. 
At inference time, ThinkRepair retrieves high-quality reasoning process exemplars from these chains of thought as few-shot prompts. This approach repairs more bugs than neural machine translation (NMT) and simple LLM baselines~\cite{Yin2024ISSTA}. 
These results suggest that incorporating reasoning processes into prompts may mitigate the reliance on surface-level patterns and improve the accuracy of log statement generation.




In this study, we introduce ThinkLog, an end-to-end approach to log statement generation that explicitly models task-specific reasoning processes.
ThinkLog fills a key gap in existing end-to-end methods, which do not explicitly model why developers write that log statement.
The key idea of ThinkLog is to construct a \textit{reasoning pool} that stores reasoning process examples generated in advance by an LLM.
These examples are derived from real-world software and explain why a log statement is inserted at a particular location, assigned a specific severity level, and includes certain message content.
Given a target method, ThinkLog retrieves similar code snippets with log statements from this pool, along with their corresponding reasoning process examples.
ThinkLog uses the retrieved code snippets, log statements, and reasoning process examples as few-shot prompts to guide the LLM in determining where to insert log statements, which severity levels to assign, and what messages to include.

We empirically evaluate ThinkLog using Java methods collected from GitHub to assess the impact of reasoning retrieved from the reasoning pool.
Our evaluation is conducted on a large-scale dataset derived from 1,465 public GitHub repositories, comprising 80,798 training instances, 10,098 validation instances, and 9,619 test instances, where each instance corresponds to a single log statement generation task constructed from a Java method.
This empirical investigation is structured around the following research questions:

\begin{description}
    \item[\textbf{RQ1:}] \RQone
    \item[\textbf{RQ2:}] \RQtwo
    \item[\textbf{RQ3:}] \RQthree
    \item[\textbf{RQ4:}] \RQfour
\end{description}

Our experiments showed that ThinkLog achieved 20.55\% perfect accuracy in log statement generation, which corresponds to a 15.4\% improvement over the best existing approach.
Using short reasoning compressed to fewer than 50 tokens led to a decline in performance.
We also found that increasing the number of examples and changing the LLM used for reasoning generation brought only limited performance gains.
In addition, ThinkLog showed stronger robustness on unseen code generated through semantics-preserving code transformations, with a smaller performance drop than existing approaches.

In summary, this study makes the following contributions:
\begin{itemize}
    \item
    We present ThinkLog, a novel end-to-end log statement generation approach centered on a reusable \textit{reasoning pool} that stores decision rationales and injects retrieved rationales into warmup and few-shot inference.
    ThinkLog is publicly available in our replication package.
    \item
    We empirically evaluate ThinkLog on a dataset of Java methods collected from public GitHub repositories. The results show that incorporating reasoning into the prompt improves log statement generation performance.
\end{itemize}

\section{Related Work} \label{sec:relatedwork}
This section reviews related work on automated log statement generation and software engineering research that uses reasoning to improve the performance of LLMs.

\subsection{Traditional Log Statement Generation} \label{subsec:traditional}

Log statement generation typically consists of three subtasks: determining log insertion positions, predicting log severity levels, and generating log messages.
As foundational work aimed at improving the accuracy of log statement generation, numerous studies have focused on automating each of these subtasks~\cite{Li2018ESE,Li2021ASE,Li2021ICSE,Ding2022SANER,Ding2023ACM}.
Li et al.~\cite{Li2021ASE} proposed a deep learning framework that estimates log insertion positions at the code block level by incorporating both syntactic and semantic information from source code.
To predict log severity levels, Li et al.~\cite{Li2021ICSE} proposed DeepLV, an approach that leverages a deep learning model.
Ding et al.~\cite{Ding2022SANER} proposed LoGenText, which generates natural language log messages using neural machine translation.

While these studies aim to automate each of the three subtasks, they do not integrate them into a single pipeline that achieves end-to-end log statement generation. Such integration is necessary to fully automate log statement generation.



\vspace{-1em}
\subsection{LLM-based End-to-End Log Statement Generation} \label{subsec:plm_llm}


LLMs have enabled end-to-end approaches that jointly handle log insertion positions, severity levels, and messages~\cite{Mastropaolo2022ICSE,Xie2024ISSTA,Xu2024ICSE}.
LANCE~\cite{Mastropaolo2022ICSE}, the first LLM-based end-to-end approach, fine-tunes T5 and achieves about 12\% end-to-end accuracy, while its extension LEONID~\cite{Mastropaolo2024JSS} combines deep learning with information retrieval but gains only marginal accuracy improvements at substantially higher computational cost.
FastLog~\cite{Xie2024ISSTA} improves insertion-position prediction by splitting the input code and predicting insertion positions at the token level, achieving about 16\% end-to-end accuracy.
SCLogger~\cite{li2024FSE} incorporates project-level static-analysis information, such as call graphs, as inter-method context in the prompt and improves log statement generation accuracy over prior LLM-based approaches.

UniLog~\cite{Xu2024ICSE} is another state-of-the-art end-to-end approach and serves as the basis for ThinkLog.
It combines few-shot prompting with a warmup strategy and achieves over 70\% accuracy for log insertion position and severity-level prediction.
In few-shot prompting, UniLog retrieves log-annotated methods similar to a target method and includes them as examples in the prompt.
In the warmup strategy, UniLog performs lightweight parameter updates with prompt-style training inputs, requiring less than 4\% of the tuning time compared to fine-tuning.

While these studies demonstrate that LLM-based end-to-end log statement generation is promising, their performance remains limited. 
For example, UniLog does not report end-to-end prediction accuracy~\cite{Xu2024ICSE}. 
LANCE and FastLog achieve end-to-end accuracies of about 12\% and 16\%, respectively~\cite{Mastropaolo2022ICSE,Xie2024ISSTA}.
This limited end-to-end accuracy suggests that current methods still struggle to reflect the purpose behind logging decisions, beyond producing plausible insertion positions, severity levels, and messages.
In fact, a systematic mapping study on logging practices reports that the literature has predominantly concentrated on where and what, leaving why relatively less investigated despite its importance for effective logging~\cite{Guang2023TSE}.
Motivated by this observation, this study aims to improve end-to-end log statement generation by addressing the missing why.


\subsection{Enhancing LLMs with Reasoning for Software Engineering Tasks}
\label{subsec:reasoning}
To improve the performance of LLMs on various software engineering tasks, prior studies have explored providing task-specific reasoning processes that guide the outputs of LLMs~\cite{Yin2024ISSTA,Yang2026ICSE}.

For automated program repair, ThinkRepair~\cite{Yin2024ISSTA} collects chains of thought generated during bug repair, stores them as a reasoning pool, and reuses selected reasoning examples at inference time.
This design improves accuracy over existing LLM-based automated program repair approaches.

For code generation, CodeThinker has been proposed~\cite{Yang2026ICSE}. 
CodeThinker collects human discussions about solving coding tasks from LeetCode and uses them as reasoning paired with corresponding code solutions. 
By fine-tuning an LLM on these reasoning-augmented samples, CodeThinker achieves higher code-generation accuracy than LLM-based baselines such as GPT-4o.





These studies suggest that providing reasoning processes tailored to software engineering tasks can enhance the performance of LLMs.
Moreover, as demonstrated by ThinkRepair, collecting reasoning processes in advance and reusing them at inference time can improve the performance without incurring significant additional costs, such as preparing static analysis information for each target method, as required by SCLogger.
Motivated by these findings, this study proposes ThinkLog, which builds a reasoning pool by collecting reasoning specific to log statement generation and utilizes it at inference time. This design aims to improve the accuracy of end-to-end log statement generation.



\section{ThinkLog} \label{sec:methodology}
\begin{figure*}[!tbp]
  \centering
  \includegraphics[width=0.86\textwidth]{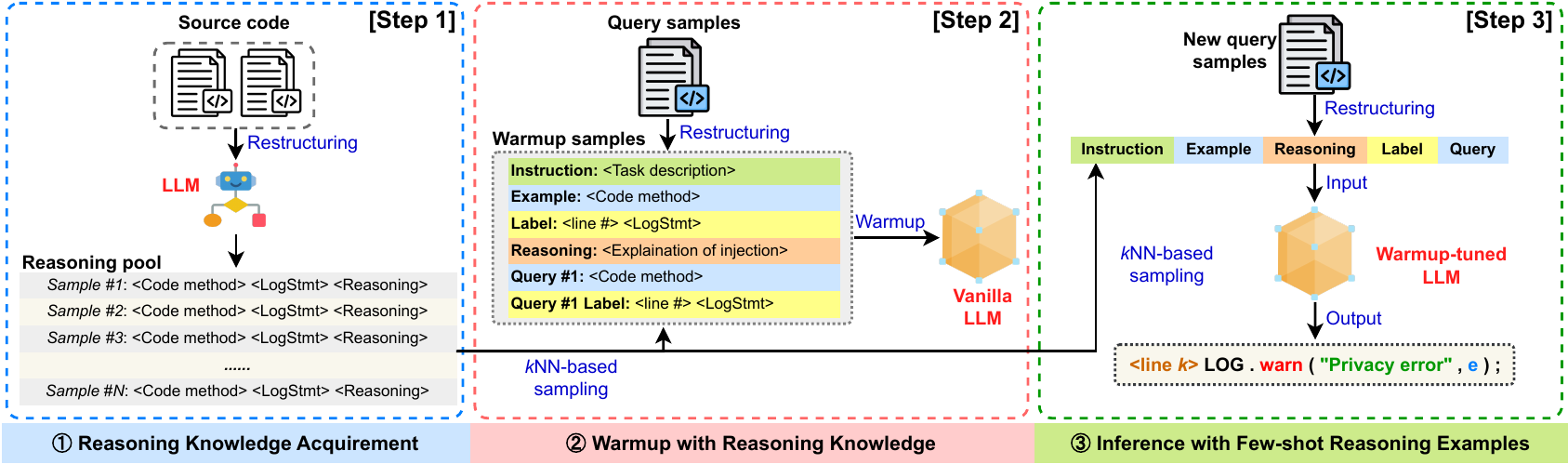}
  \vspace{-1em}
  \caption{Overview of ThinkLog}
  \label{flow}
\end{figure*}

The generation accuracy of existing LLM based end-to-end log statement generation methods remains limited.
This study proposes ThinkLog, which builds on UniLog~\cite{Xu2024ICSE}, an LLM based end-to-end log statement generation method, and is centered on a reusable reasoning pool that improves the capability of LLMs for log statement generation.
Figure~\ref{flow} presents an overview of ThinkLog.
ThinkLog consists of the following three steps.


\noindent\textbf{Step 1: Reasoning Knowledge Acquirement.}
This step builds a \textit{reasoning pool} that stores decision rationales for log statement insertion.

\noindent\textbf{Step 2: Warmup with Reasoning Knowledge.}
This step aligns the LLM with reasoning-augmented prompts through lightweight warmup.

\noindent\textbf{Step 3: Inference with Few Shot Reasoning Examples.}
This step generates log statements using few-shot examples retrieved from the \textit{reasoning pool}.

\subsection{Reasoning Knowledge Acquirement} \label{subsec:reasoning_acquirement}
In this step, we construct the \textit{reasoning pool}, a foundational component of ThinkLog. We define ``reasoning'' as the explicit articulation of decision rationales for the three critical aspects of log generation: insertion location, log level, and message content. 
Because these elements are highly interdependent, models often struggle to produce coherent combinations when relying solely on local code context. 
To address this, ThinkLog utilizes the \textit{reasoning pool} as a guiding signal during both the warmup (Step~2) and few-shot inference (Step~3) phases, ensuring that the model adheres to consistent and logical decision criteria.

To generate reasoning, we use training-split instances and provide the LLM with both the source method and its corresponding ground-truth log statement, and instruct it to articulate the rationales behind the insertion position, assigned level, and message content.
For example, the generated reasoning explains why a log statement should be inserted after a particular control-flow branch, why a specific log level is appropriate for the runtime event, and what state or contextual information should be included in the message.
Each \textit{reasoning pool} entry is stored as a triplet consisting of the source method, target log statement, and generated reasoning.
The resulting pool is constructed once before warmup and inference and reused to construct prompts in both stages.

\subsection{Warmup with Reasoning Knowledge} \label{subsec:warmup}

The main purpose of this step is to align the model with explicit decision rationales for log statement generation.
In this step, we follow the warmup strategy whose effectiveness was demonstrated in UniLog, and we further attach reasoning about log statement insertion to each example so that the model also learns decision rationales for log statement generation.

To make warmup effective with a small number of examples, we follow UniLog and retrieve examples that are similar to the target method.
Specifically, we use the \textit{reasoning pool} as the retrieval corpus and select the top-$k$ reasoning-augmented examples based on cosine similarity to the target method.
This design provides examples with similar code semantics and syntax to the query samples. 
It enables the LLM to learn where and how to log by learning the decision rationales from the top-$k$ examples with their corresponding reasoning.

Each warmup prompt consists of three components: (i) an instruction defining the task, (ii) reasoning-augmented examples that include a method, the ground truth log statement, and its reasoning, and (iii) a target method for log statement insertion with its ground truth labels.
Crucially, each training instance is augmented with reasoning traces. This enables the model, during the warmup phase, to align its inference process with explicit decision rationales.

\subsection{Inference with Few-shot Reasoning Examples}
\label{subsec:inference}
In this step, we employ a reasoning-augmented few-shot prompt to guide the warmed-up LLM in generating log statements. 
Incorporating reasoning traces at inference time reduces the reliance of models on superficial similarities among the few-shot examples, thereby promoting more consistent and robust generation.

To construct the prompt, we use the same cosine-similarity retrieval procedure as in warmup to select the top-$k$ reasoning-augmented examples from the \textit{reasoning pool}. 
These examples provide references for determining the insertion point, log level, and message content. 
The target method is then appended as the query sample, prompting the LLM to synthesize the final log statement end-to-end.

\section{Experimental Setup} \label{sec:experimentalsetup}
This section describes the experimental design used to evaluate ThinkLog, including the dataset, baselines, models, and evaluation metrics.

\subsection{Dataset} \label{subsec:dataset}
To evaluate ThinkLog and ensure a fair comparison with state-of-the-art baselines, we utilize the established Java dataset from prior studies~\cite{Mastropaolo2022ICSE,Xie2024ISSTA,Xu2024ICSE}.

\subsubsection{Dataset Construction and Preprocessing} \label{subsec:DataPreprocessing}

In this study, we use the FastLog dataset~\cite{Xie2024ISSTA} for both training and evaluation.
This dataset is identical to the LANCE dataset~\cite{Mastropaolo2022ICSE}, a widely adopted benchmark in log statement generation research~\cite{Xu2024ICSE,Xie2024ISSTA}.
It consists of 1,465 non-forked Java repositories that declare a dependency on Apache Log4j~\cite{log4j} and satisfy the original activity criteria: at least 500 commits, 10 contributors, and more than 10 GitHub stars, reducing toy or inactive projects.

Following LANCE~\cite{Mastropaolo2022ICSE} and UniLog~\cite{Xu2024ICSE}, each sample consists of a source method and one target log statement removed from that method.
For methods containing multiple log statements, we remove one target log at a time while keeping the remaining logs as context; thus, a method with $n$ log statements produces $n$ samples.

We follow UniLog's preprocessing pipeline~\cite{Xu2024ICSE} with additional filtering and standardization.
We discard methods without log statements or with syntax errors, standardize the remaining code using \texttt{google-java-format}\footnote{\url{https://github.com/google/google-java-format/}}, and extract log statements via regular expressions.
Each extracted log is annotated with a \texttt{<line\#>} tag indicating its original position, and each source method is flattened into a single sequence by prefixing every line with its corresponding tag.

We follow LANCE's original method-level 80/10/10 split protocol, resulting in \textbf{80,798 training samples, 10,098 validation samples, and 9,619 test samples}.
Thus, samples derived from the same original method are assigned to the same split and do not appear across training, validation, and test sets.
The training set is used as the retrieval corpus and the source of reasoning-pool entries, the validation set is used for warmup instances and instruction-tuning monitoring, and the test set is reserved for final evaluation.

\subsubsection{Code Transformation for Unseen Data Evaluation(RQ4)} \label{subsec:CodeTransformation}

Although LLMs have demonstrated remarkable capabilities across diverse tasks~\cite{hou2024large}, the ubiquity of large-scale web scraping~\cite{Gao2020arXiv} introduces the risk of data leakage, where models memorize benchmarks rather than learning generalizable patterns~\cite{li2024TSE}. 
To eliminate this confounder and evaluate robustness on truly unseen data, we employ lightweight and semantics-preserving transformations. 
These perturbations produce novel test instances that retain the original logic but differ in surface form, thereby preventing retrieval-based shortcuts.

To fairly assess generalization to unseen code, we construct an unseen-code split by applying semantics-preserving code transformations to the preprocessed dataset from Section~\ref{subsec:DataPreprocessing}, following prior work~\cite{li2024TSE}.
The toolkit rewrites source code at the Abstract Syntax Tree (AST) level while preserving program behavior and readability; its transformations include equivalence rewriting of conditional expressions, rewriting constants and assignments, and converting loop structures between \texttt{for} and \texttt{while}.
For example, Condition-Dup adds logically neutral clauses such as \texttt{\&\& true} or \texttt{\textbar\textbar{} false}, and Condition-Swap swaps symmetrical operands in conditional expressions, producing syntactically different but semantically equivalent code.

\subsection{Baselines} \label{subsec:baseline}
To evaluate the effectiveness of ThinkLog for end-to-end log statement generation, we adopt UniLog~\cite{Xu2024ICSE} and FastLog~\cite{Xie2024ISSTA} as state-of-the-art baselines.


\noindent\textbf{UniLog}~\cite{Xu2024ICSE} is an LLM-based baseline that uses warmup and few-shot prompting without reasoning. Following prior work~\cite{Shu2025TOSEM}, we reproduce UniLog with GPT-3.5-Turbo because its original Codex backbone is no longer available.



\noindent\textbf{FastLog}~\cite{Xie2024ISSTA} is a two-stage baseline that predicts the insertion position before generating the log statement. We use the official implementation with the default PLBART~\cite{PLABART} configuration.


We exclude LANCE~\cite{Mastropaolo2022ICSE} and LEONID~\cite{Mastropaolo2024JSS} because they are superseded by FastLog and UniLog, or offer only marginal gains with significant overhead~\cite{Shu2025TOSEM}. SCLogger~\cite{li2024FSE} is also excluded because its project-level static analysis requirement is incompatible with our method-level dataset.

\subsection{Models} \label{subsec:models}
We use different LLMs for reasoning generation and for warmup/inference.

\paragraph{Reasoning Generation}  
To generate reasoning for log statement insertion, we use three representative LLMs and compare their impact on ThinkLog in RQ3:
\begin{itemize}
\item \textbf{DeepSeek-R1}\footnote{\url{https://api-docs.deepseek.com/news/news250528}}: We use \texttt{DeepSeek-R1-0528} via the DeepSeek API. This is the default reasoning generator for RQ1, RQ2, and RQ4, and is also used for reasoning summarization in RQ2.
\item \textbf{GPT-3.5-Turbo}\footnote{\url{https://platform.openai.com/docs/models/gpt-3.5-turbo}}: We use \texttt{gpt-3.5-turbo-0125} via the OpenAI API as a general-purpose reasoning generator and to align with the UniLog reproduction backbone.
\item \textbf{Code Llama}~\cite{Codellama}: We use \texttt{Code Llama-Instruct (13B)} from Hugging Face as a code-oriented local model, which is the largest model available in our laboratory GPU environment~\footnote{NVIDIA RTX 6000 Ada Generation (48GB).}.
\end{itemize}

These models represent reasoning-oriented, general-purpose, and code-oriented categories; our goal is not an exhaustive model comparison.

\paragraph{Warmup and Inference}  
To ensure a fair comparison with UniLog, we use the same backbone model, \texttt{gpt-3.5-turbo-0125}, in both the warmup and inference phases.
We warm up the model for 5 epochs and set the decoding temperature to $0$, following UniLog.
Regarding the number of shots, we use a 1-shot setting by default: ThinkLog retrieves the single most similar example (top-1) from the reasoning pool.
We use this 1-shot setting for all RQs except RQ2, where we conduct an ablation study on the number of shots.

\subsection{Evaluation Metrics} \label{subsec:evaluationmetrics}
To ensure comparability with prior work~\cite{Xie2024ISSTA,Xu2024ICSE}, we use four exact-match accuracy metrics:
\textit{Position Accuracy}, \textit{Level Accuracy}, and \textit{Message Accuracy} measure whether the predicted insertion position, log level, and message exactly match the ground truth, respectively.
\textit{All Accuracy} measures whether all three elements match simultaneously.

Since Message Accuracy cannot capture paraphrases, we also report sentence-level BLEU~\cite{BLEU} with smoothing on a 0--100 scale and sentence-level ROUGE-L~\cite{ROUGE} on a 0--100 scale, following existing log-message generation work~\cite{Xie2024ISSTA,Shu2025TOSEM}.

\section{Results} \label{sec:results}

\subsection{RQ1: \RQone} \label{subsec:rq1}

\noindent
\textit{Approach:}
To address RQ1, we compare the effectiveness of ThinkLog with the state-of-the-art end-to-end logging statement generation methods, UniLog and FastLog.
We evaluate on the dataset constructed by the procedure in Section~\ref{subsec:DataPreprocessing}.
For this RQ, ThinkLog employs DeepSeek-R1 to generate reasoning, and leverages GPT-3.5-Turbo for warmup with reasoning and for inference with few-shot reasoning examples, as described in Sections~\ref{subsec:warmup} and~\ref{subsec:inference}.
We then report \textit{Position}, \textit{Level}, \textit{Message}, and \textit{All Accuracy}, together with \textit{BLEU} and \textit{ROUGE} as message quality metrics, as defined in Section~\ref{subsec:evaluationmetrics}.
To evaluate the online inference cost-effectiveness of ThinkLog, we compare its inference cost with UniLog, which is an LLM-based method similar to ThinkLog.
The inference cost is computed based on the official OpenAI API pricing.\footnote{\url{https://platform.openai.com/docs/pricing}. We use the listed prices for gpt-3.5-turbo-0125: \$0.50 per 1 million input tokens and \$1.50 per 1 million output tokens.}
It is important to note that UniLog and ThinkLog use the same model, i.e., GPT-3.5-Turbo.

\smallskip
\noindent
\textit{Results:}
Table~\ref{tab:results} shows the performance of FastLog, UniLog, and ThinkLog across six evaluation metrics: \textit{Position}, \textit{Level}, \textit{Message}, and \textit{All Accuracy}, as well as \textit{BLEU} and \textit{ROUGE} for generated log messages.
The numbers in parentheses for ThinkLog indicate the relative improvement over the stronger baseline for each metric.
\begin{table*}[!tbp]
\centering
\caption{Performance comparison of log generation methods (RQ1)}
\label{tab:results}
\vspace{-0.5em}
\scalebox{0.82}{
\begin{tabular}{lcccccc}
\toprule
\multirow{2}{*}{\textbf{Methods}} &
\multicolumn{1}{c}{\textbf{Position}} &
\multicolumn{1}{c}{\textbf{Level}} &
\multicolumn{3}{c}{\textbf{Message}} &
\multicolumn{1}{c}{\textbf{All}} \\
\cmidrule(lr){2-2} \cmidrule(lr){3-3} \cmidrule(lr){4-6} \cmidrule(lr){7-7}
 & \textbf{Acc.}
 & \textbf{Acc.}
 & \textbf{Acc.}
 & \textbf{BLEU}
 & \textbf{ROUGE}
 & \textbf{Acc} \\
\midrule
FastLog  & \textbf{68.94} & 32.12 & 1.37 & 9.91 & 28.04 & 1.00 \\
UniLog   & 57.93 & \underline{70.96} & \underline{22.01} & \underline{22.28} & \underline{40.96} & \underline{17.80} \\
ThinkLog & \underline{60.31} (\textcolor{red}{-12.5\%}) &
           \textbf{71.04} (\textcolor{green!45!black}{+0.1\%}) &
           \textbf{23.56} (\textcolor{green!45!black}{+6.6\%}) &
           \textbf{24.4} (\textcolor{green!45!black}{+9.5\%}) &
           \textbf{42.84} (\textcolor{green!45!black}{+4.6\%}) &
           \textbf{20.55} (\textcolor{green!45!black}{+15.4\%}) \\
\bottomrule
\end{tabular}
}
\vspace{-0.8em}
\end{table*}

\textbf{ThinkLog improved end-to-end log statement generation, especially complete correctness and message quality.}
ThinkLog outperformed UniLog across all metrics and achieved 20.55\% in \textit{All Accuracy}, representing a 15.4\% improvement over UniLog (17.80\%). ThinkLog also achieved the highest \textit{Message Accuracy} of 23.56\%, as well as the highest BLEU and ROUGE scores. These results suggest that incorporating reasoning helps the model generate log statements that better align with the intended insertion position, severity level, and message content.

\textbf{ThinkLog improves the end-to-end objective while preserving competitive component-level performance.}
Although FastLog achieved the highest \textit{Position Accuracy}, this advantage did not translate into better end-to-end log statement generation. This result is consistent with FastLog's two-stage design, which explicitly predicts insertion positions before generating log statements. In contrast, ThinkLog substantially outperformed FastLog in \textit{Level Accuracy} and achieved a level accuracy comparable to UniLog, suggesting that GPT-based prompting better captures method-level context for severity selection. More importantly, ThinkLog achieved the strongest message-related performance, which directly contributed to its best \textit{All Accuracy}.

\textbf{ThinkLog achieved these improvements with lower online inference cost than UniLog.}
In our experiments, the inference cost of UniLog was about \$9.8, whereas that of ThinkLog was about \$4.9, representing about a 50\% reduction. This comparison excludes the one-time offline cost of constructing the reasoning pool, which is amortized when the pool is reused.

\subsection{RQ2: \RQtwo} \label{subsec:rq2}
\noindent
\textit{Approach:}
We disentangle the effects of the amount of reasoning and the number of few-shot examples on ThinkLog performance.
For both ablations, we keep all other settings identical to RQ1 and evaluate ThinkLog using the same metrics as in RQ1.

\noindent\textbf{1) Varying the length of reasoning (information ablation).}
The original reasoning obtained in Section~\ref{subsec:reasoning_acquirement} contains 328 tokens on average.
We summarize it with an LLM under three length conditions ($\leq 50$, $\leq 100$, and $\leq 200$ tokens), and replace the original reasoning with the summarized reasoning in both Warmup (Step~2 of Figure~\ref{flow}) and Inference (Step~3 of Figure~\ref{flow}).


\noindent\textbf{2) Varying the number of shots (shot ablation).}
For the shot ablation, we fix the reasoning length to 100 tokens or fewer and vary the number of examples to 1, 3, and 5, following UniLog~\cite{Xu2024ICSE}.

\smallskip

\noindent
\textit{Results:}
Tables~\ref{tab:reasoning_length} and~\ref{tab:reasoning_shot} report the results for varying reasoning length and the number of examples, respectively.

\textbf{The length of reasoning particularly affects \textit{Message Accuracy}, and a sufficient amount of information is required to achieve high performance.}
When the reasoning length was 100 tokens or fewer, ThinkLog maintained performance comparable to the original reasoning (average 328 tokens). In contrast, reasoning of 50 tokens or fewer caused a clear drop in message-related metrics, including \textit{Message Accuracy}, BLEU, and ROUGE. Our manual inspection suggests that overly short summaries often preserve only the high-level logging intent while omitting concrete message-design cues, such as what event or state should be logged and how it should be phrased. This loss of concrete guidance likely makes it harder for the model to generate messages that match the reference.

\begin{table}[!tbp]
\centering
\caption{Effect of reasoning length (RQ2)}
\label{tab:reasoning_length}
\vspace{-0.5em}
\scalebox{0.76}{
\begin{tabular}{lcccccc}
\toprule
\multirow{2}{*}{\textbf{Reasoning}} &
\multicolumn{1}{c}{\textbf{Position}} &
\multicolumn{1}{c}{\textbf{Level}} &
\multicolumn{3}{c}{\textbf{Message}} &
\multicolumn{1}{c}{\textbf{All}} \\
\cmidrule(lr){2-2} \cmidrule(lr){3-3} \cmidrule(lr){4-6} \cmidrule(lr){7-7}
\textbf{length}
 & \textbf{Acc}
 & \textbf{Acc}
 & \textbf{Acc}
 & \textbf{BLEU}
 & \textbf{ROUGE}
 & \textbf{Acc} \\
\midrule
50  & 60.04 & 68.59 & 19.53 & 19.17 & 31.78 & 16.43 \\
100 & 61.62 & 71.84 & 23.74 & 24.67 & 43.28 & 20.33 \\
200 & 60.59 & 70.99 & 24.12 & 24.99 & 43.74 & 20.52 \\
Original & 60.31 & 71.04 & 23.56 & 24.4 & 42.84 & 20.55 \\
\bottomrule
\end{tabular}
}
\vspace{-0.8em}
\end{table}

\begin{table}[!tbp]
\centering
\caption{Effect of prompt example numbers (RQ2)}
\label{tab:reasoning_shot}
\vspace{-0.5em}
\scalebox{0.76}{
\begin{tabular}{lcccccc}
\toprule
\multirow{2}{*}{\textbf{Example Num.}} &
\multicolumn{1}{c}{\textbf{Position}} &
\multicolumn{1}{c}{\textbf{Level}} &
\multicolumn{3}{c}{\textbf{Message}} &
\multicolumn{1}{c}{\textbf{All}} \\
\cmidrule(lr){2-2} \cmidrule(lr){3-3} \cmidrule(lr){4-6} \cmidrule(lr){7-7}
 & \textbf{Acc}
 & \textbf{Acc}
 & \textbf{Acc}
 & \textbf{BLEU}
 & \textbf{ROUGE}
 & \textbf{Acc} \\
\midrule
1  & 61.62 & 71.84 & 23.74 & 24.67 & 43.28 & 20.33 \\
3  & 61.08 & 72.19 & 25.05 & 25.89 & 44.67 & 21.42 \\
5  & 62.00 & 72.19 & 25.20 & 26.12 & 45.04 & 21.78 \\
\bottomrule
\end{tabular}
}
\vspace{-0.8em}
\end{table}




\textbf{The performance improvement gained by increasing the number of shots is limited.}
The differences in accuracy caused by increasing the number of examples were minimal, indicating that adding more examples does not lead to further improvement.
For example, \textit{All Accuracy} was 20.33\% with 1-shot and increased only slightly to 21.42\% with 3-shot and 21.78\% with 5-shot, resulting in a difference of less than 2\%.
This result is consistent with UniLog, which reports only modest improvements from 1- to 3- to 5-shot prompts~\cite{Xu2024ICSE}.
On the other hand, since the number of input tokens increases proportionally with the number of shots, the inference cost increased while the performance gain remained limited.
These results indicate that the benefit of reasoning can be obtained even with a 1-shot prompt when using sufficiently informative reasoning.




\subsection{RQ3: \RQthree} \label{subsec:rq3}

\noindent
\textit{Approach:}
To address RQ3, we construct reasoning pools using DeepSeek-R1, GPT-3.5-Turbo, and CodeLlama, as described in Section~\ref{subsec:models}, while keeping all other settings identical to RQ1.


\smallskip
\noindent
\textit{Results:}
Table~\ref{tab:reasoning_model_comparison} compares ThinkLog under the three reasoning generation models with UniLog.
\begin{table}[!tbp]
\centering
\caption{Accuracy comparison across different reasoning generation models (RQ3)}
\label{tab:reasoning_model_comparison}
\vspace{-0.5em}
\scalebox{0.68}{
\begin{tabular}{lcccccc}
\toprule
\multirow{2}{*}{\textbf{Model}} &
\multicolumn{1}{c}{\textbf{Position}} &
\multicolumn{1}{c}{\textbf{Level}} &
\multicolumn{3}{c}{\textbf{Message}} &
\multicolumn{1}{c}{\textbf{All}} \\
\cmidrule(lr){2-2} \cmidrule(lr){3-3} \cmidrule(lr){4-6} \cmidrule(lr){7-7}
 & \textbf{Acc}
 & \textbf{Acc}
 & \textbf{Acc}
 & \textbf{BLEU}
 & \textbf{ROUGE}
 & \textbf{Acc} \\
\midrule
DeepSeek-R1       & 60.31 & 71.04 & 23.56 & 24.4 & 42.84 & 20.55 \\
GPT-3.5-turbo     & 60.99 & 70.76 & 23.12 & 24.22 & 42.52 & 19.76 \\
Codellama-13b-Instruct     & 60.67 & 71.59 & 23.06 & 23.99 & 42.6 & 19.92 \\
\midrule
UniLog (No Reasoning) & 57.93 & 70.96 & 22.01 & 22.28 & 40.96 & 17.80 \\
\bottomrule
\end{tabular}
}
\vspace{-0.8em}
\end{table}

\textbf{The choice of the model used for generating reasoning has a minor impact on the final performance of ThinkLog.}
DeepSeek-R1 achieved the highest \textit{All Accuracy}, but the differences among the three reasoning generation models were less than 1\%, indicating that the final performance is relatively insensitive to the choice of reasoning generator. CodeLlama also achieved comparable performance while being runnable locally in our setting, making it a practical option. Across all three models, ThinkLog consistently outperformed UniLog, reaffirming that incorporating reasoning contributes to improved log generation performance.

\subsection{RQ4: How well does ThinkLog generalize to unseen code?} \label{subsec:rq4}

\noindent
\textit{Approach:}
To address RQ4, following the procedure in Section~\ref{subsec:CodeTransformation}, we evaluate ThinkLog on transformed code while keeping all other settings identical to RQ1.
We report the difference from the original test set to quantify degradation, and compare only UniLog and ThinkLog because FastLog's extremely low \textit{All Accuracy} in RQ1 makes degradation analysis unstable.

\smallskip
\noindent
\textit{Results:}
Table~\ref{tab:unseen_code} presents the results on the transformed dataset, together with the performance degradation rate ($\Delta$) relative to the original test data.
\begin{table*}[!tbp]
\centering
\caption{Performance of log statement generation on unseen code (RQ4)}
\label{tab:unseen_code}
\vspace{-0.5em}
\scalebox{0.78}{
\begin{tabular}{lcccccccccccc}
\toprule
\multirow{2}{*}{\textbf{Method}} & 
\multicolumn{2}{c}{\textbf{Position}} & 
\multicolumn{2}{c}{\textbf{Level}} & 
\multicolumn{6}{c}{\textbf{Message}} & 
\multicolumn{2}{c}{\textbf{All}} \\
\cmidrule(lr){2-3} \cmidrule(lr){4-5} \cmidrule(lr){6-11} \cmidrule(lr){12-13}
 & \textbf{Acc} & $\boldsymbol{\Delta}$ 
 & \textbf{Acc} & $\boldsymbol{\Delta}$ 
 & \textbf{Acc} & $\boldsymbol{\Delta}$ 
 & \textbf{BLEU} & $\boldsymbol{\Delta}$ 
 & \textbf{ROUGE} & $\boldsymbol{\Delta}$ 
 & \textbf{Acc} & $\boldsymbol{\Delta}$ \\
\midrule
UniLog   & 63.60 & 9.8\%↑ 
         & 67.50 & 4.9\%↓ 
         & 16.14 & 26.7\%↓ 
         & 17.85 & 19.9\%↓ 
         & 34.73 & 15.2\%↓ 
         & 12.43 & 30.2\%↓ \\
ThinkLog & 69.80 & 15.7\%↑ 
         & 72.29 & 1.8\%↑ 
         & 20.11 & 14.6\%↓ 
         & 23.19 & 5.0\%↓ 
         & 41.4 & 3.4\%↓ 
         & 16.30 & 20.7\%↓ \\
\bottomrule
\end{tabular}
}
\vspace{-0.8em}
\end{table*}

\textbf{ThinkLog exhibited strong generalization capability even on unseen code.}
For the message-related metrics (\textit{Message Accuracy}, \textit{BLEU}, and \textit{ROUGE}) and \textit{All Accuracy}, both methods exhibited performance degradation, but ThinkLog showed a smaller decline than UniLog (Table~\ref{tab:unseen_code}).
In particular, \textit{All Accuracy} dropped from 17.80\% to 12.43\% for UniLog (30.2\%) and from 20.55\% to 16.30\% for ThinkLog (20.7\%).
These results suggest that reasoning helps ThinkLog maintain stronger generalization performance on unseen code.
Interestingly, both methods achieved higher \textit{Position Accuracy} on transformed code, possibly because transformations such as \textsc{LocalVar} and \textsc{Assignment} alter the salience ranking of candidate insertion lines while keeping the ground-truth line relatively high.

\section{Discussion} \label{sec:discussion}

ThinkLog improves log statement generation by retrieving decision rationales from the reasoning pool, but the mechanism behind this improvement requires further analysis.
To investigate this point, the first author manually inspected 100 randomly sampled cases (out of 599) where UniLog failed but ThinkLog succeeded; the second author then reviewed the sampled cases and the derived insights.

Based on this inspection, we found that successful reasoning often explicitly covered three points:
\begin{description}
  \item[Position] Where the log statement should be inserted and why.
  \item[Level] What logging objective the statement serves and which severity level fits it.
  \item[Message] What information the message should convey and to whom.
\end{description}
These points provide practical guidance for the model when it must jointly decide the insertion position, severity level, and message.

\begin{figure}[!tbp]
  \centering
  \includegraphics[width=0.6\columnwidth]{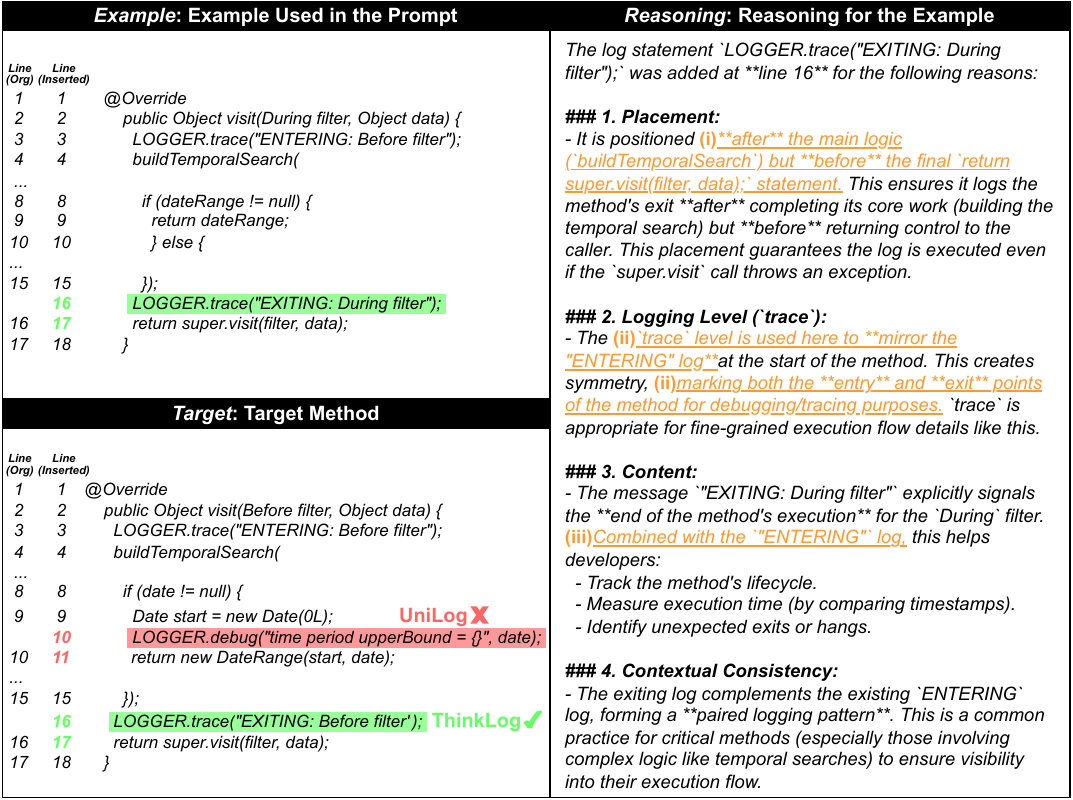}
  \vspace{-1.2em}
  \caption{(A) few-shot example, (B) its reasoning, and (C) a target method for log insertion}
  \label{fig:example_query}
\end{figure}

\smallskip
\noindent
\textit{Example.}
Figure~\ref{fig:example_query} shows a case where UniLog fails but ThinkLog succeeds.
The example and target share a similar control-flow pattern, but UniLog selects an in-branch debug log, whereas ThinkLog follows reasoning that states the insertion boundary, the tracing-oriented level, and the \textit{EXITING}-style message, leading it to the ground-truth output.

\textbf{As a practical guideline, we recommend that the reasoning included in the prompt explicitly covers position, level, and message decisions.}
The reasoning pool is the mechanism that stores such rationales and makes them reusable during warmup and inference.

\section{Threats to Validity} \label{sec:threats}

\noindent \textbf{Construct validity:}
We exclude SCLogger because it requires project-level static analysis, whereas our dataset is provided at the method level. Future work should extend the dataset with project-level information to enable comparisons with such methods.

\noindent \textbf{Internal validity:}
A potential threat is the reproduction of UniLog, whose implementation is not publicly available. We reimplemented it based on the original paper~\cite{Xu2024ICSE} and include our implementation in the replication package. Another threat is possible data leakage from the pretraining corpus of \texttt{GPT-3.5-Turbo}; to mitigate this concern, RQ4 evaluates ThinkLog on transformed unseen code and shows that it still outperforms the baselines.

\noindent \textbf{External validity:}
Our evaluation focuses on Java projects and three reasoning-generation LLMs. Although Java is widely used in prior log-generation studies~\cite{Mastropaolo2022ICSE,Xie2024ISSTA,Xu2024ICSE} and ThinkLog is not language-specific, results may differ for other languages or LLMs. The offline reasoning pool may also become stale when the target codebase or logging practice changes. Future work should evaluate ThinkLog across broader programming languages, model families, and evolving codebases.



\section{Conclusion} \label{sec:conclusion}

This paper introduces ThinkLog, an end-to-end log statement generation approach centered on a reusable \textit{reasoning pool} that stores reasoning generated in advance and retrieves rationales for input prompts.
Our evaluation on 9,619 Java methods from GitHub showed that ThinkLog achieves 20.55\% log statement generation accuracy, a 15.4\% improvement over UniLog.
It also limited performance degradation to 20.7\% on unseen code, compared with 30.2\% for UniLog, while keeping the online inference cost in USD at approximately 50\% of UniLog's cost.
\textbf{This study empirically demonstrates that retrieving decision rationales from a reasoning pool can improve accuracy, generalization, and online inference cost efficiency in end-to-end log statement generation.}

\section{Data Availability} \label{sec:availability}
Our replication package can be accessed at \url{https://doi.org/10.5281/zenodo.18308001}.

\subsubsection*{Acknowledgments.}
We gratefully acknowledge the financial support of: (1) JSPS for the KAKENHI grants (JP24K02921, JP25K03100, JP25K22845, JP26H02500, JP26K21198); (2) Japan Science and Technology Agency (JST) as part of Adopting Sustainable Partnerships for Innovative Research Ecosystem (ASPIRE), Grant Number JPMJAP2415, and (3) the Kayamori Foundation of Informational Science Advancement for supporting Tao Xiao, and (4) the Inamori Research Institute for Science for supporting Yasutaka Kamei via the InaRIS Fellowship.

\subsubsection*{Disclosure of Interests.}
All co-authors have seen and agree with the contents of the manuscript, and there is no financial interest to report.

\end{document}